\documentclass{aa}
\usepackage{graphicx}
\usepackage{txfonts}
\usepackage{color}
\usepackage{xspace}
\usepackage{siunitx}
\usepackage{amsmath, amsfonts}
\usepackage{amssymb}
\usepackage{xcolor}
\usepackage{dashbox}
\usepackage{framed}
\usepackage{lipsum}
\usepackage{placeins}
\usepackage{stfloats}
\usepackage{picture}
\usepackage{multirow}
\usepackage{tikz}
\usepackage{hyperref}

\hypersetup{
        colorlinks=true, 
        breaklinks=true,
        linkcolor=blue, 
        citecolor=blue, 
        filecolor=blue, 
        urlcolor=blue,
        unicode=false, 
        pdftoolbar=true, 
        pdfmenubar=true, 
        pdffitwindow=false, 
        pdfstartview={Fit}, 
        pdftitle={}, 
        pdfauthor={Alena Rottensteiner}, 
        pdfsubject={},
        pdfcreator={Alena Rottensteiner}, 
        pdfkeywords={},
        pdfnewwindow=true, 
        pdfdisplaydoctitle=true 
}

\defcitealias{Rottensteiner_2026}{R26}
\defcitealias{Kounkel_2020}{K20}
\defcitealias{Grossschedl_2021}{G21}
\defcitealias{Swiggum_2021}{S21}
\defcitealias{Foley_2023}{F23}

\begin{document}

\title{The star formation history of Orion is structured and episodic}

\author{Alena Rottensteiner\inst{1} \and
Jo\~ao Alves\inst{1,2}\and
Stefan Meingast\inst{1}\and 
Sebastian Hutschenreuter\inst{1}\and
Núria Miret-Roig\inst{3}\and
Josefa Großschedl\inst{4}\and 
Cameren Swiggum\inst{5}\and
Sebastian Ratzenböck\inst{1,2,5} \and  
Monika G. Petr-Gotzens\inst{6}
 }

\institute{Department of Astrophysics, University of Vienna, Türkenschanzstrasse 17, 1180 Wien, Austria 
\\ \email{alena.kristina.rottensteiner@univie.ac.at}
 \and
University of Vienna, Research Network Data Science at Uni Vienna, Kolingasse 14-16, 1090 Wien, Austria
\and
Faculty of Physics, Martí i Franquès, 1-11, 08028 Barcelona, Spain
\and
Astronomical Institute of the Czech Academy of Sciences, Bocní II 1401, 141 31 Prague 4, Czech Republic
\and
Center for Astrophysics | Harvard \& Smithsonian, 60 Garden St., Cambridge, MA 02138, USA
\and 
European Southern Observatory, Karl-Schwarzschild-Straße 2, 85748 Garching bei München, Germany
}

\date{Received 30 September 2026 / Accepted xxx}

\abstract{
The propagation of star formation throughout a molecular cloud complex can tell us much about its physical drivers and the importance of environmental versus intrinsic processes.
We reconstruct the star formation history (SFH) of the Orion star-forming complex using the highest-resolution age map currently available. It comprises 47 co-spatial and co-moving groups identified by the \texttt{SigMA} clustering algorithm, with homogeneously derived ages and group memberships.
We find four periods of increased star formation, separated by $\sim5$~Myr each. The two periods of highest increase, at 6.4 and 11.3 Myr, can be statistically localized, together hold ca.~59\% of all stars, and are robust against resampling and changes to the age fitting. The two older, smaller increases at 16.2 and 21.4~Myr can be identified in the age distribution, but cannot be shown to be robust at this point. The $\sim5$~Myr spacing between the episodes of increased star formation is the same as found for the nearby Scorpius-Centaurus (Sco-Cen) star-forming region. We find no methodological bias that could produce the SFH structure. We identify spatio-temporally connected structures, including three cluster-chain candidates. We conclude that star formation in Orion seems to occur on the same timescales but with a different pattern than in Sco-Cen. The cause of the $\sim$5 Myr separations remains a mystery, and its detection in two distinct star-forming regions needs further investigation.
}
\keywords{Methods: data analysis -- statistical -- Stars: pre-main sequence -- open clusters and associations: individual: Orion}
\maketitle
\nolinenumbers  

\section{Introduction}
\label{sec:Introduction}

The star formation history (SFH) of a star-forming region, and/or OB association, encodes a wealth of information about the drivers of star formation. Such regions and associations are now known to be highly substructured in space, velocity, and time \citep[e.g.,][]{Jerabkova_2019, Kerr_2021, Posch_2025}. Their substructure is inherent and connected, and can provide vital information about the primordial state of the star-forming region \citep{Wright_2020}. One of the most important questions is whether star formation is an intrinsically chaotic process or follows reconstructible patterns at the currently resolvable group structure and $\sim$Myr timescale. Moreover, if any patterns exist, are they intrinsic to the formation process, or governed by environmental influences, such as triggering \citep[e.g.,][]{Elmegreen_1977}? Answering these questions is crucial for developing accurate star formation models, interpreting observations, and determining whether there is a connection between the SFH and the likelihood of forming gravitationally bound or unbound clusters \citep{Ratzenboeck_2023b}. 

Distinguishing the loose, mostly unbound substructures of star-forming complexes or OB associations with sufficiently high resolution only became feasible thanks to \emph{Gaia} \citep{Gaia}. While observations indicate that star formation is not a stationary process and highlight the role of stellar feedback \citep[e.g.,][]{Grossschedl_2021}, there is no consensus yet on whether star formation propagates in an ordered or chaotic fashion. A review by \citet{Wright_2022} concludes that there is little to no evidence of coherent patterns in star formation. Indeed, only a more recent investigation of the Scorpius-Centaurus (Sco-Cen) star-forming complex found global, coherent spatio-temporal structures \citep{Ratzenboeck_2023b}. This finding includes several episodes of increased star formation rate, separated by $\sim5$~Myr. A similarly episodic SFH was reported for the Cepheus Flare region \citep{Wang_2025}.

Here, we investigate the SFH of the best-studied nearby star-forming region and producer of an OB association and numerous clusters: the Orion complex. We show that its formation history is differently structured than that of Sco-Cen, but that the same $\sim5$~Myr spacing between star formation peaks appears.

\section{Dataset and method}
\label{sec:Data}

We use the Orion cluster census of \citet{Rottensteiner_2026} (\citetalias{Rottensteiner_2026}) and also adopt their cluster names (Table~\ref{tab:chains}). It was derived with the \texttt{SigMA} clustering algorithm \citep{Ratzenboeck_2023a} from \emph{Gaia} DR3 \citep{Gaia_DR3_2023} data and comprises 47 clusters\footnote{``Cluster'' in this context refers to a statistical density enhancement over a background, e.g., the field star density. We do not claim that any of the described structures are bounded.} aged 1.5--25~Myr, with 11,996 reliable members. \citetalias{Rottensteiner_2026} determined isochronal cluster ages with the method described in \cite{Ratzenboeck_2023b}, using a PARSEC isochrone grid \citep{Bressan_2012}. The cluster ages and upper/lower limits are the mode and 1$\sigma$ uncertainty intervals of the MCMC posterior distribution. If a bound was smaller than the grid spacing, \citetalias{Rottensteiner_2026} quotes the cell width as a conservative limit.

To analyze the age distribution, we compute a kernel density estimate (KDE), where each cluster is represented by a two-piece normal kernel $\mathcal{N}_2(m, \sigma_{\text{lo}}, \sigma_{\text{hi}})$ centered on its age $m$, with $\sigma_{\rm lo}$ and $\sigma_{\rm hi}$ the distances to its lower and upper $1\sigma$ bounds. The sub-Myr uncertainties of \citetalias{Rottensteiner_2026} for the youngest clusters are unrealistic, given systematics such as differential reddening, cross-contamination, and differences among isochrone models.
Therefore, we introduce a uniform age-uncertainty floor of $\sigma_{\rm floor}=1$~Myr on each side, which affects 23 clusters, all younger than $10$~Myr. We calibrate $\sigma_{\rm floor}$ by bootstrapping each cluster's members 5 times and refitting the ages using the same settings as in \citetalias{Rottensteiner_2026}. For the 24 clusters unaffected by the floor, the \citetalias{Rottensteiner_2026} uncertainties match the bootstrap scatter (median ratio 0.98). For the other 23, they are too narrow by a median factor of 1.27, and the floor brings their median back to unity. For completeness, Fig.~\ref{fig:age-kde-raw} shows the age distribution without the uncertainty floor.

\section{Results}
\label{sec:Results}

\begin{figure}[t]
    \centering
    \includegraphics[width=\linewidth]{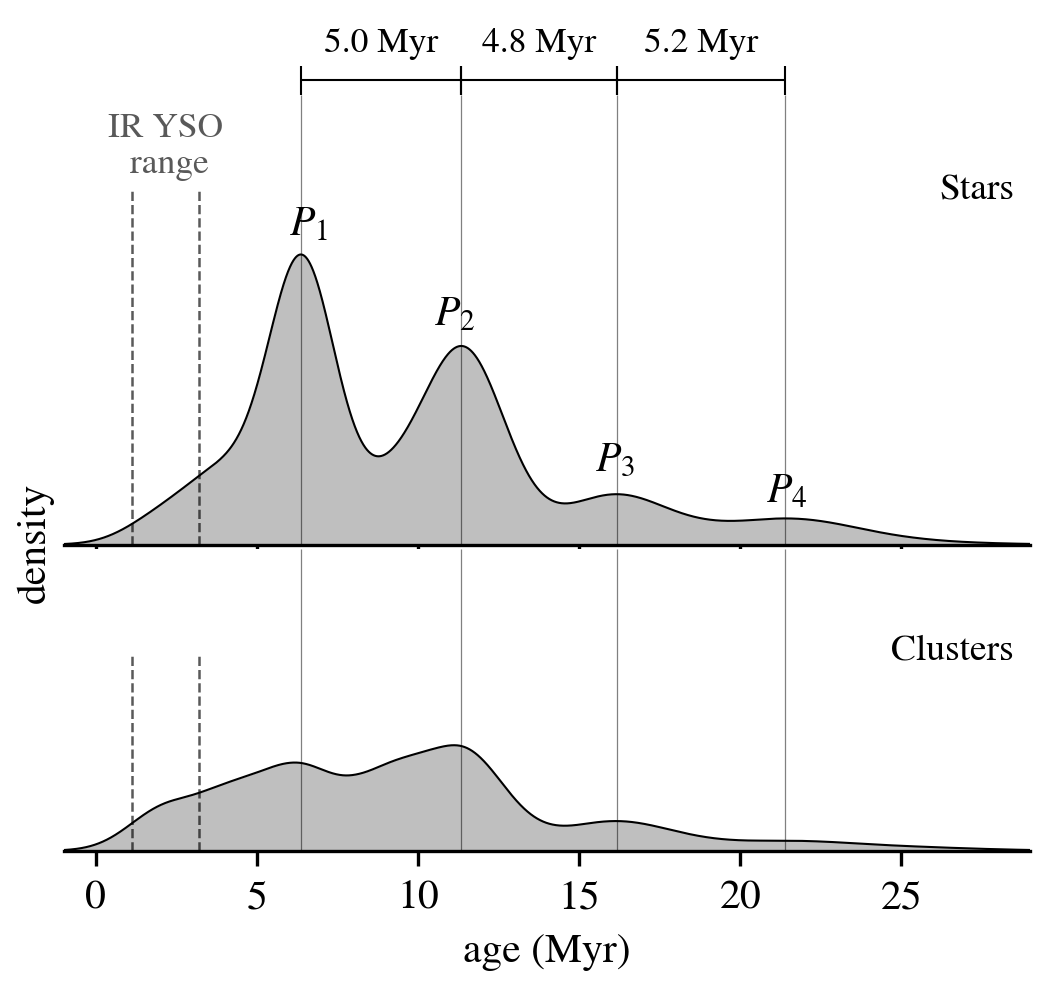}
    \caption{SFH of Orion, as the KDE of the 47 cluster ages, weighted by the number of members (\emph{top}), or unweighted (\emph{bottom}). The four identified peaks and separations are marked. Dashed lines show where embedded YSOs may add another peak between 1.1--3.2~Myr in the star panel.}
    \label{fig:age-kde}
\end{figure}

Figure~\ref{fig:age-kde} shows the age distribution for the Orion clusters and stars. Since ages were determined per cluster, the top panel is essentially a weighted version of the bottom, but star and cluster formation are not linearly correlated, as different stellar masses were formed at different times. It therefore makes sense to look at both distributions \citep{Ratzenboeck_2023b}. Orion's SFH is not smooth but instead exhibits peaks and dips, which are more apparent in the star panel but also evident in the 47-data-point cluster panel. They are also visible without using a KDE (Fig.~\ref{fig:formation-timeline}). 
We identify four peaks of star formation, which are separated by $\simeq5$~Myr each (Table~\ref{tab:peaks}). The number of stars per peak increases towards younger ages, as seen in the top panel. Both panels agree on the peak positions, with the biggest deviation at the smallest peak of $\simeq 0.4$~Myr. $P_1$ and $P_2$ have a comparable number of clusters $N_c$, but the two panels disagree on which peak comprises more. This is because the boundary between the peaks is well-defined in the star panel at $8.8$~Myr, but only a shallow dip at $7.8$~Myr for the clusters. Three cluster ages fall within this difference and shift peaks between the panels. 

The ISF(ONC) cluster contributes 2197 stars (37\%) to $P_1$. However, \citetalias{Rottensteiner_2026} could not split it into the foreground cluster NGC~1980 \citep{Alves_2012, Chen_2020} and the Orion Nebula Cluster (ONC), and its age (6.5~Myr) does not represent the ONC's estimated $\sim2$~Myr \citep{Megeath_2016}. We refitted only members within 10--30\arcmin{} of $\theta^1$~Ori to try and isolate the ONC, but all ages remained $>5.7$~Myr. This indicates that the \citetalias{Rottensteiner_2026} cluster is mainly composed of NGC~1980 stars, not ONC members, and they are not misassigned to $P_1$. Using the Young stellar objects (YSO) catalog of \citet{2025Roquette}, we find that 1854 YSOs have no \emph{Gaia} DR3 match within 2\arcsec, indicating they are very young and still embedded. 20\% lie within 20\arcmin{} of $\theta^1$~Ori, suggesting that already the input of \citetalias{Rottensteiner_2026} may miss many ONC stars. The embedded YSOs could potentially add another peak between 1.1--3.2~Myr in Fig.~\ref{fig:age-kde} or enhance the tentative shoulder seen for the cluster panel in this region (Appendix~\ref{appendix:biases-SigMA}).

\begin{table}[h!]
    \centering
    \caption{Characterization of the SFH peaks, with age ranges calculated from 5 bootstrappings of the cluster ages and $k/5$ marking peaks recovered in only $k$ refits. $N_c$ is the cluster number between adjacent minima.}
 \begin{tabular*}{\columnwidth}{@{\extracolsep{\fill}}r|ccc|ccc@{}}
\hline \hline
 & age & range & $N_c$ & age & range & $N_c$ \\
 & (Myr) & (Myr) & & (Myr) & (Myr) & \\
\hline
& \multicolumn{3}{c}{\emph{Stars}} & \multicolumn{3}{c}{\emph{Clusters}} \\
\hline
$P_1$ & 6.4 & 6.3--6.7 & 22 & 6.2 & 4.9--6.8 & 19 \\
$P_2$ &11.3 & 10.7--11.6 & 17 & 11.1 & 4/5 & 20 \\
$P_3$ &16.2 & 1/5 & 5 & 16.1 & 2/5 & 5 \\
$P_4$ &21.4 & 3/5 & 3 & 21.0 & 1/5 & 3 \\
\hline 
\end{tabular*}
\label{tab:peaks}
\end{table}

\section{Discussion}
\label{sec:Discussion}

The star formation history of Orion (Fig.~\ref{fig:age-kde}) is not smooth, but peaked. Four identified maxima are separated by a remarkably even spacing of about 5~Myr -- the same regularity that was found for Sco-Cen \citep{Ratzenboeck_2023b}. Here, we analyze the reliability of these features before discussing the SFH of Orion.

\subsection{The peaks are not methodological artifacts}
\label{sec:Discussion-method}

Since the Orion and Sco-Cen results are based on the same method, we first test whether the SFH peaks could have methodological origins. We identify three possible sources of bias: the clustering algorithm, the age-fitting, and the KDE bandwidth (Appendix~\ref{appendix:biases}). We find that none of them is responsible for the structure. The clustering only uses astrometry and is independent of the photometry-based ages. Accounting for embedded YSOs does not change the observed peaks, but it could add another, younger peak. It could follow the 5~Myr spacing, but to verify it, we would need accurate age estimates for these sources. Correcting for member loss due to cluster dispersal at the old end would only increase the heights of $P_3$ and $P_4$, without affecting cluster ages. Regarding the age-fitting, the grid and priors are both uniform in $\log t$, which cannot produce evenly spaced features in linear age. Refitting all cluster ages while varying the fit functions, isochrone model, magnitude range, and CMDs recovers $P_1$ and $P_2$ in all variants with the necessary grid resolution. Their absolute positions can shift by up to 1.7 and 1.3 Myr, but their separation remains stable to within 0.6 Myr. $P_3$ and $P_4$ are not recovered as reliably, likely due to their smaller height and fewer members. Lastly, we show that the four described peaks remain stable in the KDE plot up to the theoretical resolution limit near 1.6~Myr under variation of $\sigma_{\rm floor} \in [0,2]$~Myr.

\subsection{Statistical reliability and spatial evidence for the peaks}
  \label{sec:Discussion-statistics-physical}

A KDE maximum is not by itself proof of a distinct underlying population. Therefore, we investigate the statistical reliability of the results (Appendix~\ref{appendix:stats}). Permitting two modes, an excess-of-mass test \citep{Mueller_1991} localizes $P_1$ and $P_2$ between 6.2--6.5 and 10.9--12.3~Myr, respectively. Together they hold $\sim59\%$ of all member stars, and 16 of all 47 clusters. To evaluate the significance of these modes, we perform a calibrated dip test \citep{Hartigan_1985, Cheng_1998}, which rejects unimodality for the star distribution at the 10\% level ($p = 0.064$). However, it does not reject unimodality for the clusters ($p = 0.77$), and a Gaussian mixture model (GMM) prefers a single component in both panels. The cluster and GMM results can be explained by the shallow minimum between $P_1$ and $P_2$ in the cluster distribution, and the large combined peak uncertainties.
Bootstrap refits of the cluster ages recover $P_1$ and $P_2$ in all five runs for the star panel, but $P_3$ and $P_4$ are only reproduced in one and three runs, respectively (Table~\ref{tab:peaks}). Lastly, the regularity of the $\sim5$~Myr spacings of the peaks cannot be reproduced by smooth star formation histories. We conclude that $P_1$ and $P_2$ are localized modes of the star distribution at the 10\% significance level, and robust to resampling. $P_3$ and $P_4$ can be described from Fig.~\ref{fig:age-kde}, but we cannot prove their significance. Whether this means they are not real, or can be explained by missing members in the census, will be the topic of future work.

As an independent test, we investigate whether the clusters of each peak are also spatially correlated. Using 3D positions, we find that $68\%$ of the clusters have their nearest neighbor in the same peak, compared to $34\%$ for random peak labels. The correlation is significant ($p = 0.0002$) and also holds for sky positions ($p = 0.008$). Thus, clusters born at the same time are likely also neighbors. This can also be seen visually in Figs.~\ref{fig:SFH-spatial} and \ref{fig:3D-scatter}, with the exception of $P_3$, which shows no spatial coherence.

\begin{figure*}[h]
    \centering
    \includegraphics[width=\linewidth]{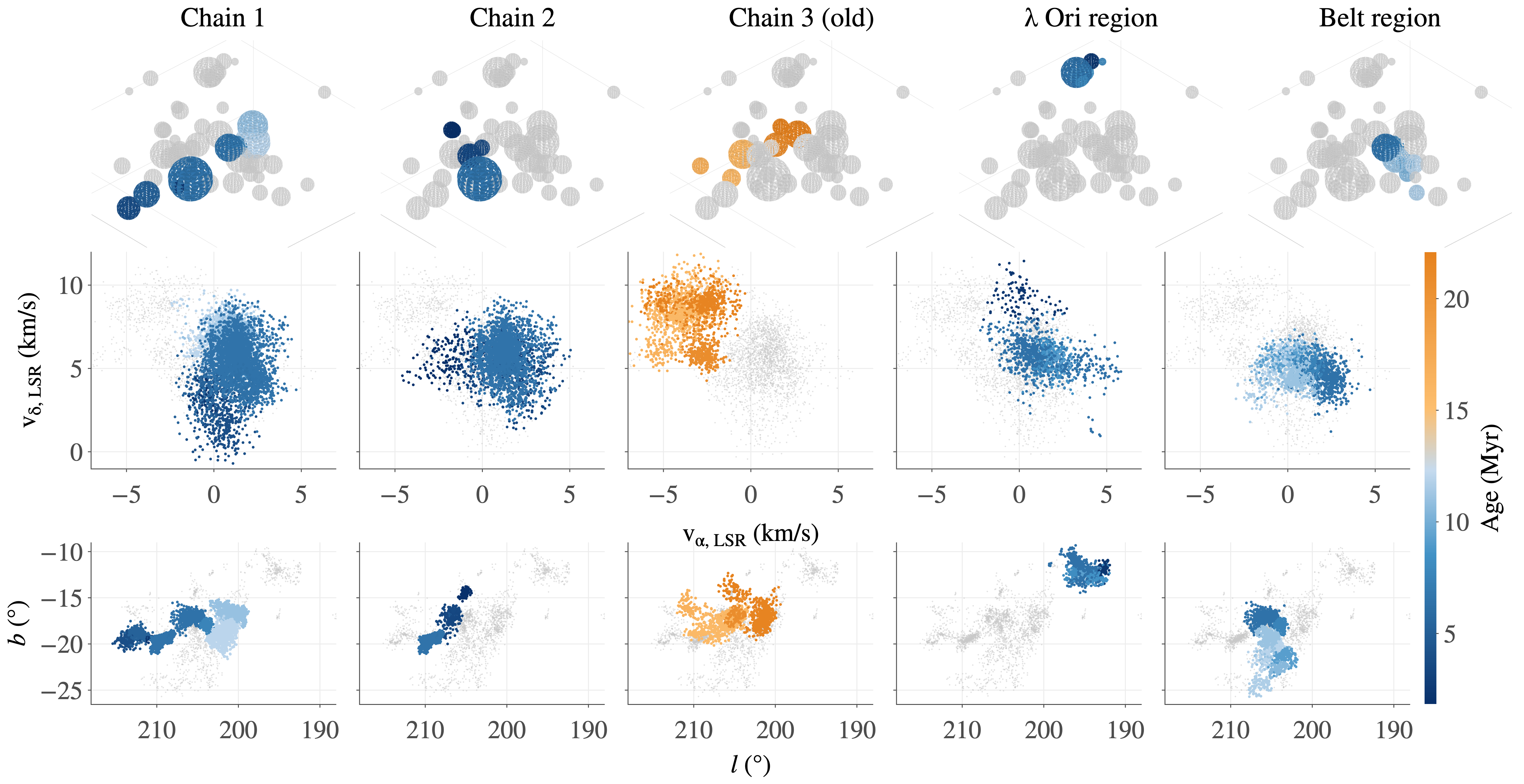}
\caption{Spatio-temporal structures identified in Orion, color-coded by age.
  \emph{Top row:} Clusters in 3D heliocentric Galactic Cartesian coordinates, shown as spheres with radius $\propto N^{1/3}$ at their
  median position, in the view of the top row of Fig.~\ref{fig:SFH-spatial}.
  \emph{Middle row:} Tangential velocities of the member stars in the local standard of rest (LSR).
  \emph{Bottom row:} Galactic sky positions of the member stars.}
    \label{fig:chains}
\end{figure*}

\subsection{Is star formation in Orion feedback-driven?}
\label{sec:Discussion-Cluster-chains}

In Sco-Cen, star formation propagated in an inside-out fashion throughout the complex, driven by stellar feedback \citep[e.g.,][]{Krause_2018, Ratzenboeck_2023b, Posch_2023, Posch_2025, Miret-Roig_2025}. We investigate whether the same holds for Orion, using the clusters' on-sky and 3D positions, and tangential velocities in the local standard of rest (LSR). Our goal is to probe the possible \emph{existence} of patterns, not to conduct a full analysis. We manually identify spatial and temporal correlations among the clusters, defining ``chains'' as spatially elongated groups of clusters that exhibit an age gradient, and ``regions'' as spatially correlated clusters without a clear age gradient.
We chose this approach over an automated one for two reasons: First, the significant line-of-sight (LOS) elongation of the clusters hampers the detection of spatial correlations (Fig.~\ref{fig:3D-scatter}). Second, missing embedded stars and extinction can render young cluster ages less reliable. For instance, in \citetalias{Rottensteiner_2026}, NGC~2024 is 3.5 Myr old, but at least twice as many embedded, younger members are missing from this census \citep{Levine_2006, Rottensteiner_2026a}. Therefore, we consider a structure a cluster chain candidate even if its age sequence is not strictly monotonic. Contrary to Sco-Cen, Orion does not appear to exhibit a global, inside-out cluster formation pattern, but we do find evidence of local age gradients and sequential star formation on a shorter timescale. We identify three chain candidates and two regions, involving 31 clusters (66\%) in Fig.~\ref{fig:chains} and describe them in Table~\ref{tab:chains}: 

Chain 1 (``Orion-A chain'') extends $\sim130$~pc from the Orion~A cloud to the Orion belt population (OBP) and includes two $P_2$ and seven $P_1$ clusters. We note that its central part (ONC to OBP-near) is the classical Orion OB1(a-d) association \citep{Blaauw_1964}, which has long been proposed for triggered star formation \citep{Elmegreen_1977}. Except for L1641-C and the ONC, age increases monotonically within the uncertainties, and it is the only chain whose tangential velocity ($v_{\delta, \rm LSR}$) also increases with age ($\rho = 0.93$).
Chain~2 (``Orion-B chain'') extends $\sim50$~pc backward from the ISF(ONC) towards Orion~B and contains seven $P_1$ clusters with a strictly monotonic age sequence, except for RV~Orionis. Chain 3 (``old chain'') covers $\sim$90~pc and includes all $P_4$ clusters and most of $P_3$. The two remaining $P_3$ clusters lie $>45$~pc away, which is why $P_3$ shows no coherence in Sect.~\ref{sec:Discussion-statistics-physical}. It is well separated from the younger chains in tangential velocity and may have been a source of feedback acting on them. The chain clusters are highlighted in Fig.~\ref{fig:formation-timeline} as well. The regions we identify are the $\lambda$~Ori and the Belt. 

We can contextualize these results within the proposed global formation scenarios of Orion: \citet{Kounkel_2020} proposed one, and \citet{Grossschedl_2021} (\citetalias{Grossschedl_2021}) one or more supernovae $\sim6$~Myr ago as a trigger for the ONC. \citetalias{Grossschedl_2021} also showed that the head of the Orion~A cloud, the Orion~B cloud, and several cometary clouds expand radially from the same region, and attributed the formation of the ONC and NGC~2024 to feedback-driven compression. \citet{Swiggum_2021} (\citetalias{Swiggum_2021}) found coherent stellar expansion in OBP-near and Briceño-1, and \citet{Foley_2023} (\citetalias{Foley_2023}) placed these two clusters at the center of the Orion Shell, an expanding dust cavity with the Orion~A, Orion~B, and $\lambda$~Ori clouds on its surface. We do not measure expansion, but our census resolves Orion's substructure and ages in more detail than previous studies. Our ages follow the order of \citetalias{Swiggum_2021}, with Briceño-1 older than OBP-near. Except for NGC~2024, all clusters younger than 5~Myr lie on or outside the shell surface, and all other clusters in its interior are older than 6~Myr, as predicted by the triggered star formation scenario of \citetalias{Foley_2023}. Across all 47 clusters, however, age does not correlate with distance from the expansion center of \citetalias{Swiggum_2021}, in line with \citetalias{Grossschedl_2021}, who consider a single event unlikely to explain all of Orion.
Beyond the previous works, we find that Chains~1 and 3 both start inside the Orion Shell, within 40~pc of the expansion center of \citetalias{Swiggum_2021}, and become younger outward. They are almost parallel, have a separation of only $\approx25$~pc, and aligned age gradients. This means we find a tentative spatial relation between clusters separated by more than $10$~Myr in age, connecting clusters from all four star formation peaks. Chain~3 predates the events discussed above, consistent with the suggestion of \citetalias{Swiggum_2021} that gas left over from an older epoch influenced the assembly of the younger populations.

In summary, we find a different SFH pattern, but the same timescales in Orion and Sco-Cen. Does this challenge the feedback-driven star formation established from Sco-Cen? On the contrary -- it merely reinforces the importance of different environments for star formation: Orion is more massive than Sco-Cen, it has much more residual gas, and its largest star-formation peak is the most recent one, whereas Sco-Cen's SFH peaked 15~Myr ago. As such, we expect more (ongoing) massive stellar feedback, which can both nurture star formation by compressing residual gas and triggering the birth of new generations, and be destructive, by ionizing the molecular gas and dispersing clouds \citep[e.g.,][]{Pabst_2019, Lewis_2023, Neralwar_2024}. Combined with the larger gas reservoir, this presents more opportunities for complex interactions between the stars and the gas. While the age gradients are more local in Orion than Sco-Cen, we also find connections between the positions of many of the young clusters and their ISM surroundings (Fig.~\ref{fig:end-of-chain}), which are indicative of stellar feedback \citep{Kalari_2018, Grossschedl_2021, Posch_2023, Posch_2025}. The $\sim5$~Myr gaps between star formation peaks in both Sco-Cen and Orion may even relate to the propagation speed of stellar feedback, but this topic requires further research.

\section{Conclusion}
\label{sec:Conclusion}

We investigated the star formation history of the entire Orion complex, based on the most recent cluster census and highest-resolution age map (\citetalias{Rottensteiner_2026}). Our main result is that the SFH is episodic and structured, rather than chaotic, with four increases between 6.4 and 21.4~Myr, separated by $\sim5$~Myr each.
The two more recent peaks are statistically robust. Our finding is very similar to Sco-Cen, where \citet{Ratzenboeck_2023b} also reported episodic star formation with the same spacing. However, the patterns differ between Orion and Sco-Cen: We tentatively assign 66\% of the clusters to cluster chains and locally coherent structures, which will be characterized in future work.
Our results support a complex, feedback-driven formation scenario for Orion, and its different SFH pattern compared to Sco-Cen could be due to environmental factors. The $\sim5$~Myr spacing between star formation episodes could hint at physical timescales for feedback propagation and requires further investigation.


\begin{acknowledgements}
Co-Funded by the European Union (ERC, ISM-FLOW, 101055318). JG acknowledges co-funding from the European Union, the Central Bohemian Region, and the Czech Academy of Sciences, as part of the MERIT fellowship, MSCA-COFUND Horizon Europe, Grant agreement 101081195.
Views and opinions expressed are those of the author(s) only. 
This work has made use of data from the European Space Agency (ESA) mission
\emph{Gaia} (\url{https://www.cosmos.esa.int/gaia}), processed by the \emph{Gaia}
Data Processing and Analysis Consortium (DPAC,
\url{https://www.cosmos.esa.int/web/gaia/dpac/consortium}). Funding for the DPAC
has been provided by national institutions, in particular the institutions
participating in the \emph{Gaia} Multilateral Agreement. This research has made use of the NASA/IPAC Infrared Science Archive, which is funded by the National Aeronautics and Space Administration and operated by the California Institute of Technology.
\end{acknowledgements}

\bibliography{ref}
\clearpage

\appendix
\nolinenumbers  
\section{Description of correlated structures}
\label{appendix:chains}

Table \ref{tab:chains} lists the members and properties of the structures identified in Fig.~\ref{fig:chains} and discussed in Sect.~\ref{sec:Discussion-Cluster-chains}. Cols. 1--3 describe cluster chain candidates, i.e., elongated structures with evidence of a positional and an age gradient. Cols. 4--5 describe regions, i.e., spatially connected groups of clusters with similar ages, but without a clear age gradient. Four clusters are shared by two structures: The ISF(ONC) and NGC~1977 are assigned to Chain 1, and are also the oldest clusters of Chain 2. OBP-near and OBP-near ext. are part of Chain 1, and the youngest clusters of the Belt region. The clusters are listed in order along each structure, not along their age sequence. We note that the age gradient in the Belt region primarily stems from OBP-near and OBP-near ext., whereas the other clusters do not follow this pattern.

\begin{table*}
\caption{Tentative assignment of the \citetalias{Rottensteiner_2026} Orion clusters to a cluster chain or region (Fig.~\ref{fig:chains}). Four clusters appear multiple times. $L$ is the median separation between the two end clusters, $f_1$ is the positional variance along the principal axis, and $\rho$ is the Spearman age-position correlation. As the chains were selected manually, $\rho$ only describes how ordered each structure is, and has no other statistical significance.}
\label{tab:chains}
\centering
\begin{tabular}{@{}lr@{\hspace{1.6em}}lr@{\hspace{1.6em}}lr@{\hspace{1.6em}}lr@{\hspace{1.6em}}lr@{}}
\hline\hline
\multicolumn{2}{c}{\textbf{Chain 1 (Orion-A)}} & \multicolumn{2}{c}{\textbf{Chain 2 (Orion-B)}} & \multicolumn{2}{c}{\textbf{Chain 3 (old)}} & \multicolumn{2}{c}{\textbf{$\lambda$ Ori region}} & \multicolumn{2}{c}{\textbf{Belt region}} \\
\multicolumn{2}{c}{\footnotesize $N=9$, $L=129$\,pc} & \multicolumn{2}{c}{\footnotesize $N=7$, $L=51$\,pc} & \multicolumn{2}{c}{\footnotesize $N=6$, $L=91$\,pc} & \multicolumn{2}{c}{\footnotesize $N=6$, $L=26$\,pc} & \multicolumn{2}{c}{\footnotesize $N=7$, $L=48$\,pc} \\
\multicolumn{2}{c}{\footnotesize $f_1=98$\%, $\rho=+0.88$} & \multicolumn{2}{c}{\footnotesize $f_1=92$\%, $\rho=+0.89$} & \multicolumn{2}{c}{\footnotesize $f_1=87$\%, $\rho=+0.77$} & \multicolumn{2}{c}{\footnotesize $f_1=76$\%, $\rho=+0.49$} & \multicolumn{2}{c}{\footnotesize $f_1=81$\%, $\rho=+0.75$} \\
\hline
Cluster & Age & Cluster & Age & Cluster & Age & Cluster & Age & Cluster & Age \\
\hline
L1647 & 3.9 & NGC 2071 & 1.5 & Orion A-East & 16.7 & B30-splinter & 7.9 & OBP-near & 6.4 \\
L1641S & 5.3 & NGC 2068 & 1.9 & Orion A-West & 15.8 & B30 & 2.3 & OBP-near ext. & 7.6 \\
L1641-C & 2.4 & RV Orionis & 4.5 & OBP-far & 15.9 & $\lambda$ main B & 8.5 & $\eta$ Ori & 11.1 \\
ISF(ONC) & 6.5 & $\sigma$ Ori & 3.4 & OBP-b & 20.8 & $\lambda$ Ori & 6.3 & L1616A & 9.2 \\
NGC 1977 & 6.2 & NGC 2024 & 3.5 & SO-6 & 25.0 & $\lambda$ main C & 9.2 & SO-3 & 11.9 \\
OBP-near & 6.4 & NGC 1977 & 6.2 & ASCC 20 & 21.8 & $\lambda$ main A & 8.2 & SO-1 & 10.5 \\
OBP-near ext. & 7.6 & ISF(ONC) & 6.5 &  &  &  &  & SO-2 & 11.6 \\
Briceno-1B & 12.0 &  &  &  &  &  &  &  &  \\
Briceno-1A & 11.0 &  &  &  &  &  &  &  &  \\
\hline
\end{tabular}
\end{table*}

\begin{figure*}
    \centering
    \includegraphics[width=\linewidth]{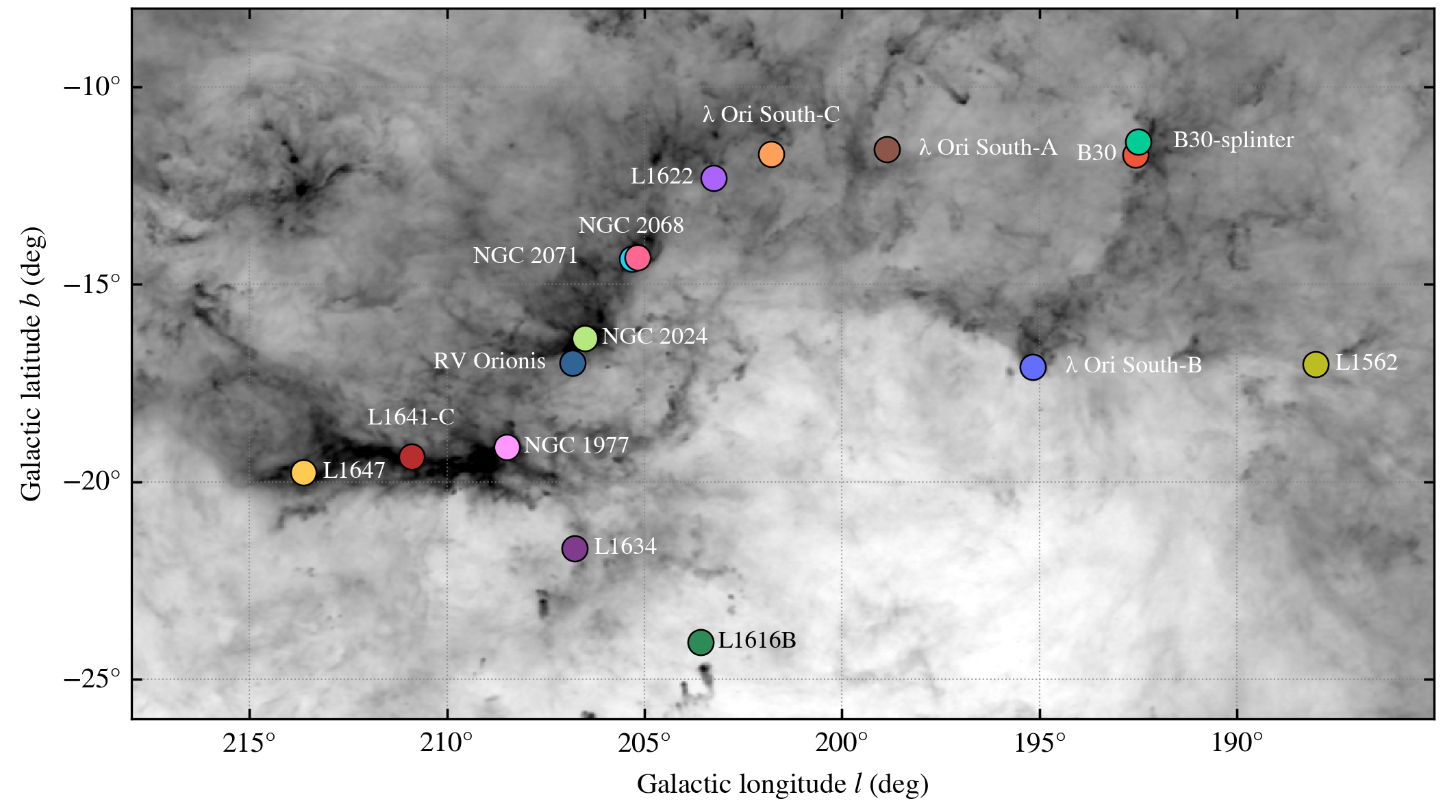}
    \caption{Young cluster positions atop the Planck 857 GHz map for the clusters that lie near prominent dust features related to star formation, such as GMCs or cometary-shaped clouds.}
    \label{fig:end-of-chain}
\end{figure*}

Figure~\ref{fig:end-of-chain} shows the median on-sky positions of 16 of the 47 Orion clusters that appear close to prominent dust features in the \emph{Planck} 857 GHz map \citep{PlanckPR3_ancillary}. The selection includes 8 of the 9 clusters with ages $<5$~Myr (exception: $\sigma$~Ori), and every selected cluster is younger than 11.3~Myr. We highlight that \texttt{SigMA} has only astrometric information, so the correlations found with the dust further validate the clustering result. The young clusters produced by the Orion~A and B clouds are well-known, but we can also relate six of the seven clusters younger than 10~Myr that are not part of a chain or region in Fig.~\ref{fig:chains} to cometary-shaped clouds, whose ionized, compressed heads point towards the groups and the fluffy tails extend away from them. These features have been interpreted as indicators of feedback-driven star formation \citep{Kalari_2018, Grossschedl_2021, Posch_2023, Posch_2025}. 

\section{Analysis of possible method biases}
\label{appendix:biases}

In this section, we evaluate potential methodological sources of bias for the presented results. We specifically investigate the type of bias that would cause a peaked distribution and/or introduce periodical effects that could cause the observed peak spacing.

\subsection{Possible biases in the clustering algorithm}
\label{appendix:biases-SigMA}

The \texttt{SigMA} clustering procedure works exclusively on the 5D astrometric solution from \emph{Gaia}, and source photometry is only introduced at the age-fitting stage. This means that the clustering step is independent of the cluster ages. However, it may still be age-selective at both edges of the SFH distribution: 

A large fraction of young, still embedded stars without \emph{Gaia} measurements could theoretically skew $P_1$, which could then turn into the shoulder of a larger, more recent star formation peak. We estimate the number of such embedded YSOs in Orion using the NEMESIS catalog \citep{2025Roquette}. It contains 15\,192 YSOs that have an infrared-based evolutionary YSO class of either 0/I, flat spectrum, II, or III thin disk. Between 1\,854 and 2\,429 YSOs have no \textit{Gaia} DR3 counterpart within a 1-2\arcsec match radius, and 49--63\% of them are from the youngest YSO classes (0/I and flat). Next, we estimate how many sources would need to be missed in the star panel of Fig.~\ref{fig:age-kde}, to change the current distribution. For $P_1$ to become the shoulder of a younger peak, at least $3\,200$ YSOs aged 4.3~Myr, 10\,000 stars at age $3.5$~Myr, and even more stars for younger ages would be needed. Stars older than 4.3~Myr would merge into $P_1$. All these numbers exceed the embedded 2\,429 YSO candidates, meaning that embedded members alone cannot skew $P_1$. However, the embedded members of \citet{2025Roquette} could still influence the peaks: Assigning them a common age between 1 and 3~Myr with the 1~Myr kernel floor would add a separate maximum at 1.1--3.2~Myr (dashed lines in Fig.~\ref{fig:age-kde}), depending on the chosen age. Spreading their ages out over this interval would produce a broader, weaker feature instead. In either case, $P_1$-$P_4$ remain unaffected. It is possible that this new peak would appear at the same $\sim5$~Myr spacing observed for $P_1$-$P_4$, or it could strengthen either or both features seen in Fig.~\ref{fig:age-kde-raw}, but we would need age estimates for all embedded YSOs to constrain their impact further. In the cluster panel of Fig.~\ref{fig:age-kde}, there is an indication of a shoulder in the peak range for the embedded YSOs already, which could be strengthened by e.g., adding the ONC as a separate population. 

At the old end, the \emph{Gaia}-based cluster census of \citetalias{Rottensteiner_2026} progressively underestimates cluster memberships because the fastest members drift beyond the density-defined cluster boundaries. By $10-30$~Myr, about a third of members lie beyond the density threshold \citep{Alves_2026}. This lowers the weight of the older clusters in the star-weighted SFH, likely contributing to the low statistical significance of $P_3$ and $P_4$. It does not, however, affect their ages and cannot cause any periodicity on its own.

\subsection{Possible biases in the age fitting method}
\label{appendix:biases-age-fitting}

The \citetalias{Rottensteiner_2026} cluster ages are based on the \textsc{Chronos} algorithm described in \cite{Ratzenboeck_2023b}. It uses a Bayesian framework to fit a user-defined fit function to stellar CMD data and compares it with a model isochrone grid. For the \citetalias{Rottensteiner_2026} ages, a PARSEC isochrones grid with a separation of 0.04~dex, the $\rm G_{\rm BP} - \rm G_{\rm RP}$ Color-Magnitude diagram (CMD), a skewed Cauchy fit function, and a fit range of $M_{\rm G} \in (-\infty, 10)$~mag were used. We test potential biases introduced during the age-fitting step by examining the architecture and operation of the fit code, and by running an explicit fit suite varying the fit parameters. 

The fitting architecture cannot imprint the structures found in Fig.~\ref{fig:age-kde}, because the PARSEC isochrone grid and sampler prior used are both uniform in logarithmic age, whereas the peaks are evenly spaced in linear age, at $5.0$, $4.8$, and $5.2$~Myr, translating to $0.25$, $0.15$, and $0.12$~dex. A mechanism that has a scale that is constant in $\log t$ is incapable of producing intervals constant in $t$.
 
To test the robustness of the results regarding the specific fit parameters, we re-derive all 47 cluster ages in five variants:
\begin{itemize}
    \item[V1] $\rm G_{} - \rm G_{\rm RP}$ CMD
    \item[V2] Least trimmed squares (LTS) fit function \citep{Rousseeuw_1984} 
    \item[V3] restricted fit range, $M_{\rm G} \in (4.5,10)$~mag
    \item[V4] BHAC15 isochrone grid \citep{Baraffe_2015} and $M_{\rm G} \in (4.5,10)$~mag
    \item[V5] BHAC15 isochrone grid and $M_{\rm G} \in (4.5,12)$~mag
\end{itemize}

The BHAC15 grid stops at an upper mass limit of $1.4\,M_{\odot}$, corresponding to a faintest magnitude of $M_{\rm G} = 4.5$~mag at 5~Myr and a brightest magnitude of $2.9$~mag at 15~Myr. To get an age-independent sample cut, we cap the magnitude range at $M_{\rm G} = 4.5$~mag. V3 is the control run for the performance of the PARSEC suite over the BHAC15 dynamic range. Because the BHAC15 grid extends to lower masses than PARSEC (0.01 vs. 0.09~$M_\odot$), we extend its dynamical range towards fainter stars in V5. The LTS fit (trim fraction $0.2$)returns no posterior, so the V2 curve uses the reference fit's kernel widths, centered on the V2 ages.
\begin{figure}
    \centering
    \includegraphics[width=\linewidth]{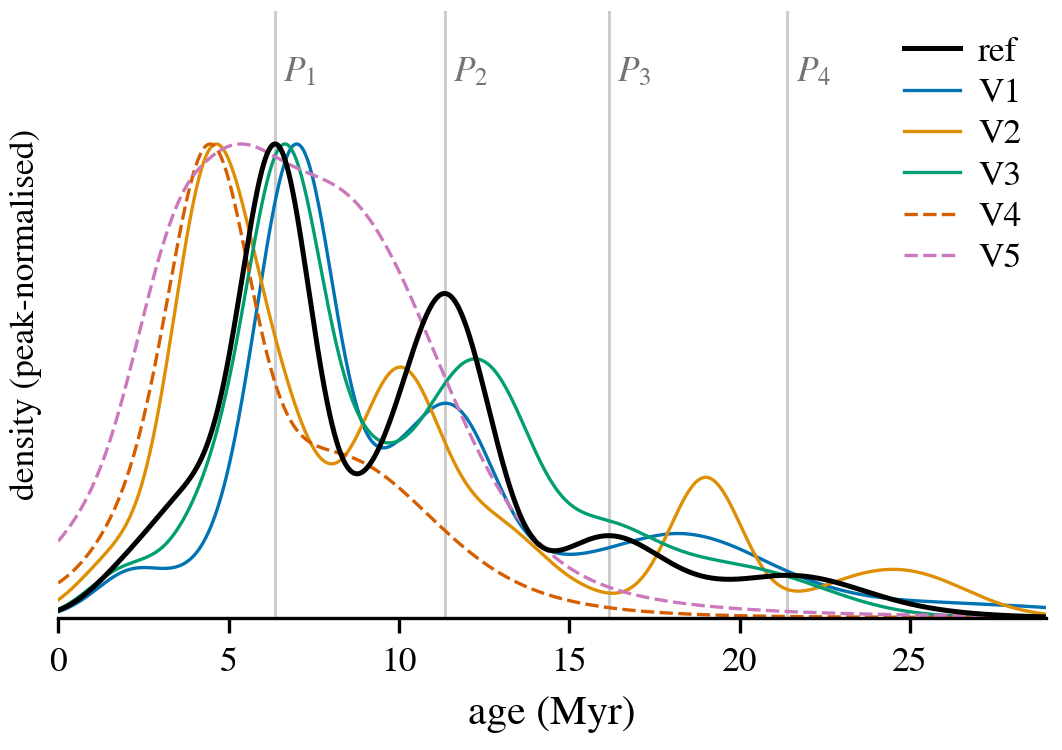}
    \caption{Star-weighted age distribution of the reference solution of Fig.~\ref{fig:age-kde} (black line), compared to five variants of the age-fitting procedure. Variants using PARSEC model isochrones are plotted as solid lines, while those using BHAC15 isochrones are shown as dashed lines.}
    \label{fig:variants}
\end{figure}
Figure~\ref{fig:variants} shows the results for the star-weighted panel. The three PARSEC variants recover $P_1$ and $P_2$ of Fig.~\ref{fig:age-kde}, separated by $4.4$ (V1), $5.4$ (V2) and $5.6$~Myr (V3), versus the $5.0$~Myr in the distribution of Sect.~\ref{sec:Results}. The absolute position of $P_1$ and $P_2$ moves by up to $1.7$ and 1.3~Myr, respectively, but the peaks shift together, and their separation deviates at most $0.6$~Myr, whereas independent motion would predict $\sim2.2$~Myr.

V1 shifts $P_1$ by $0.6$~Myr while leaving $P_2$ within $0.03$~Myr, and V3 shifts $P_1$ by $0.3$~Myr and $P_2$ by $0.9$~Myr, indicating that neither the color index nor the magnitude cut significantly affects the positions. V2 places $P_1$ at $4.6$ and $P_2$ at $10.0$~Myr and moves $P_3$ and $P_4$ to $19.0$ and $24.5$~Myr, corresponding to an age-dependent systematic shift of the peaks, while still preserving the relative spacing between $P_1$ and $P_2$, and $P_3$ and $P_4$, respectively, although the gap between $P_2$ and $P_3$ widens to 9.0~Myr. We therefore conclude that the \citetalias{Rottensteiner_2026} ages are conditional on PARSEC and the skewed-Cauchy fit at the $1-3$~Myr level, but the peak separations are affected considerably less.

The BHAC15 grid contains only twelve ages $\leq30$~Myr, spaced by $0.08-0.3$~dex, and none between 5 and 8~Myr, where $P_1$ lies in Fig.~\ref{fig:age-kde}. As the grid spacing dictates the minimum age uncertainty, the median BHAC15 age uncertainties are five times larger than for PARSEC. We also observe a compression of the age scale when using BHAC15 isochrones compared to V3: The median BHAC15 cluster ages are 38\% lower than their PARSEC counterparts, and the effect intensifies with age. V4 spans only $1.9-10.2$~Myr and contains only a single cluster older than 10~Myr, producing only a single maximum at $4.5$~Myr with a shoulder at $6.4$~Myr, which, on this compressed scale, is where $P_2$ falls. The fainter limit of V5 admits $28.5\%$ more stars and restores much of the age range ($1.8-24.7$~Myr), and a second shoulder appears near 13~Myr. However, with a median kernel width of $3.3$~Myr, dictated by the BHAC15 grid resolution, $P_1$ and $P_2$ merge into a single maximum at $5.4$~Myr. This can be explained using the merging criterion of Appendix~\ref{appendix:biases-plot}, where two maxima 5~Myr apart cannot be separated with kernels wider than 2.5~Myr, and the compressed age scale reduces their separation further. The BHAC15 variants, therefore, cannot resolve $P_1$ and $P_2$. A fair comparison would require a finer grid derived directly from the BHAC15 evolutionary tracks rather than the published isochrones, but this is beyond the scope of this test.

$P_3$ and $P_4$ of Fig.~\ref{fig:age-kde} are not recovered with the same reliability. $P_3$ appears in V1 and V2, and $P_4$ in V2, but both are displaced by $2-3$~Myr and not spaced as regularly as $P_1$ and $P_2$. This reproduces the conclusion drawn from the bootstrapping that only the two youngest episodes are robust to sample variation (Sect.~\ref{sec:Discussion-statistics-physical}). Whether the features $P_3$ and $P_4$ in Fig.~\ref{fig:age-kde} are real but unstable due to their low number statistics and high errors, or artifacts, remains unclear. Removing the member weighting and only considering the 47 cluster data points preserves the $P_1$ and $P_2$ structure across V1-V3, but the cluster panels show additional maxima -- in V2 at $4.2$~Myr and V3 at $8.6$~Myr. The unweighted separations also scatter more widely, $3.3-6.6$~Myr against $4.4-5.6$~Myr star-weighted.

Taken together, the age-fitting variants show that the absolute ages depend on the isochrone model and the fit function at the $1-3$~Myr level, but the two youngest peaks and their $\sim$5 Myr separation are recovered in every PARSEC variant. $P_3$ and $P_4$ are not robust to the fitting choices, and the BHAC15 grid is too coarse to test the structure.

\subsection{Possible biases in the presentation of the results}
\label{appendix:biases-plot}

\begin{figure}[t]
    \centering
    \includegraphics[width=\linewidth]{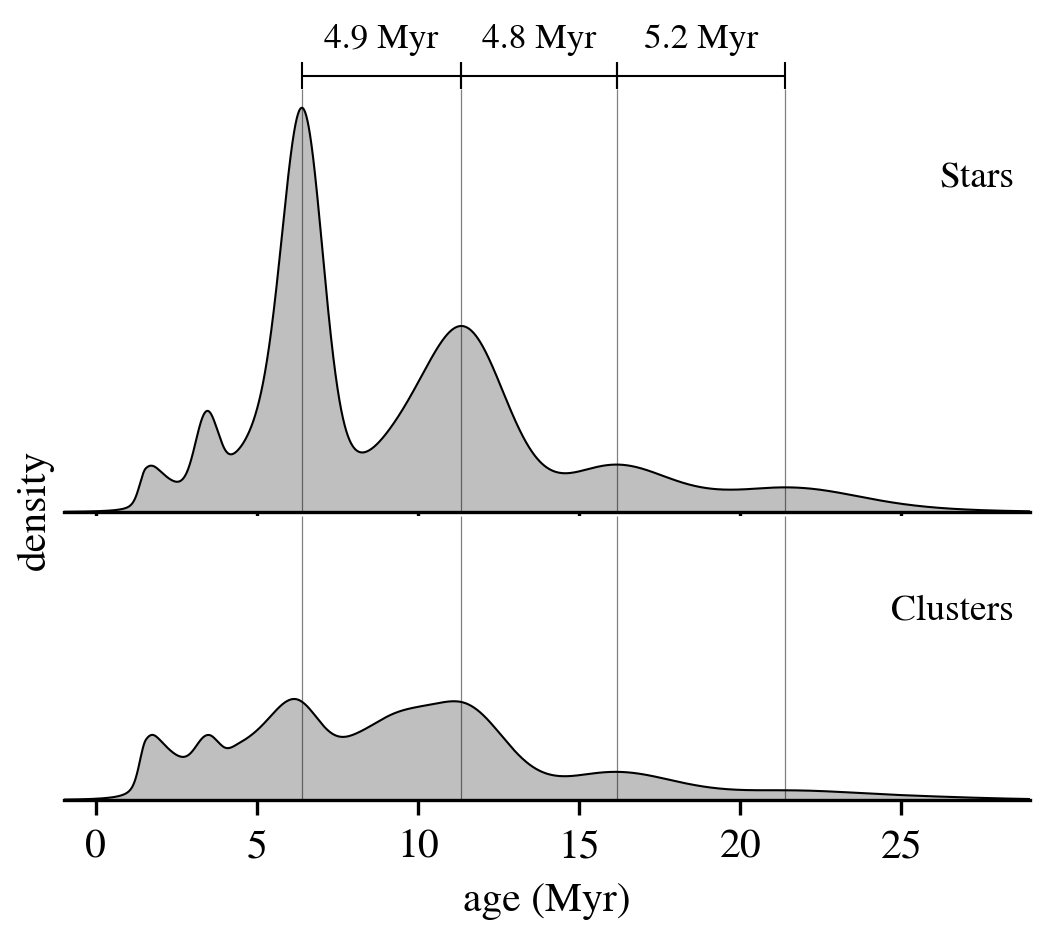}
    \caption{Same as Fig.~\ref{fig:age-kde}, but using the age uncertainties of \citetalias{Rottensteiner_2026}.}
    \label{fig:age-kde-raw}
\end{figure}

\begin{figure*}[ht!]
    \centering
    \includegraphics[width=\linewidth]{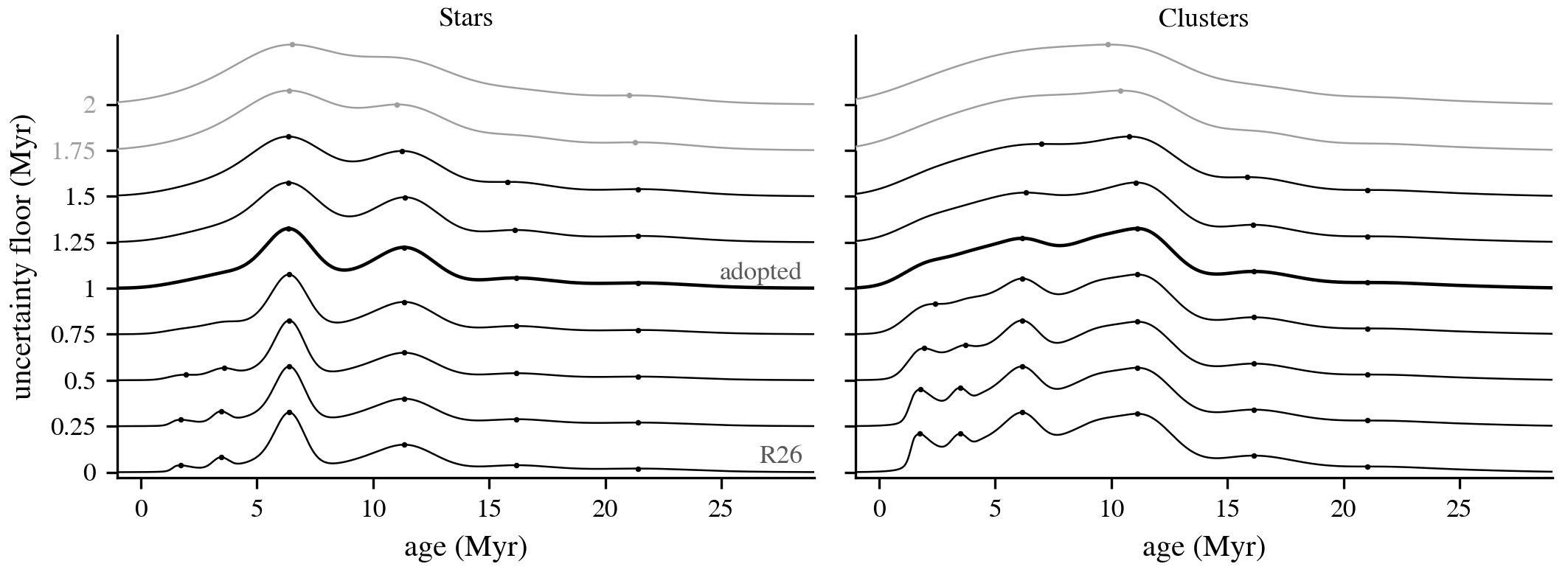}
    \caption{Age distributions of the per-star (\emph{left}) and per-cluster (\emph{right}) panels for different $\sigma_{\rm floor}$ values. The curve for the adopted value is printed in bold. Floor values above the theoretical merging limit for $P_2$ and $P_3$ in the star panel are shown in gray. Detected peaks are marked by dots.}
    \label{fig:sweep}
\end{figure*}

\begin{figure*}[ht!]
     \centering
     \includegraphics[width=\linewidth]{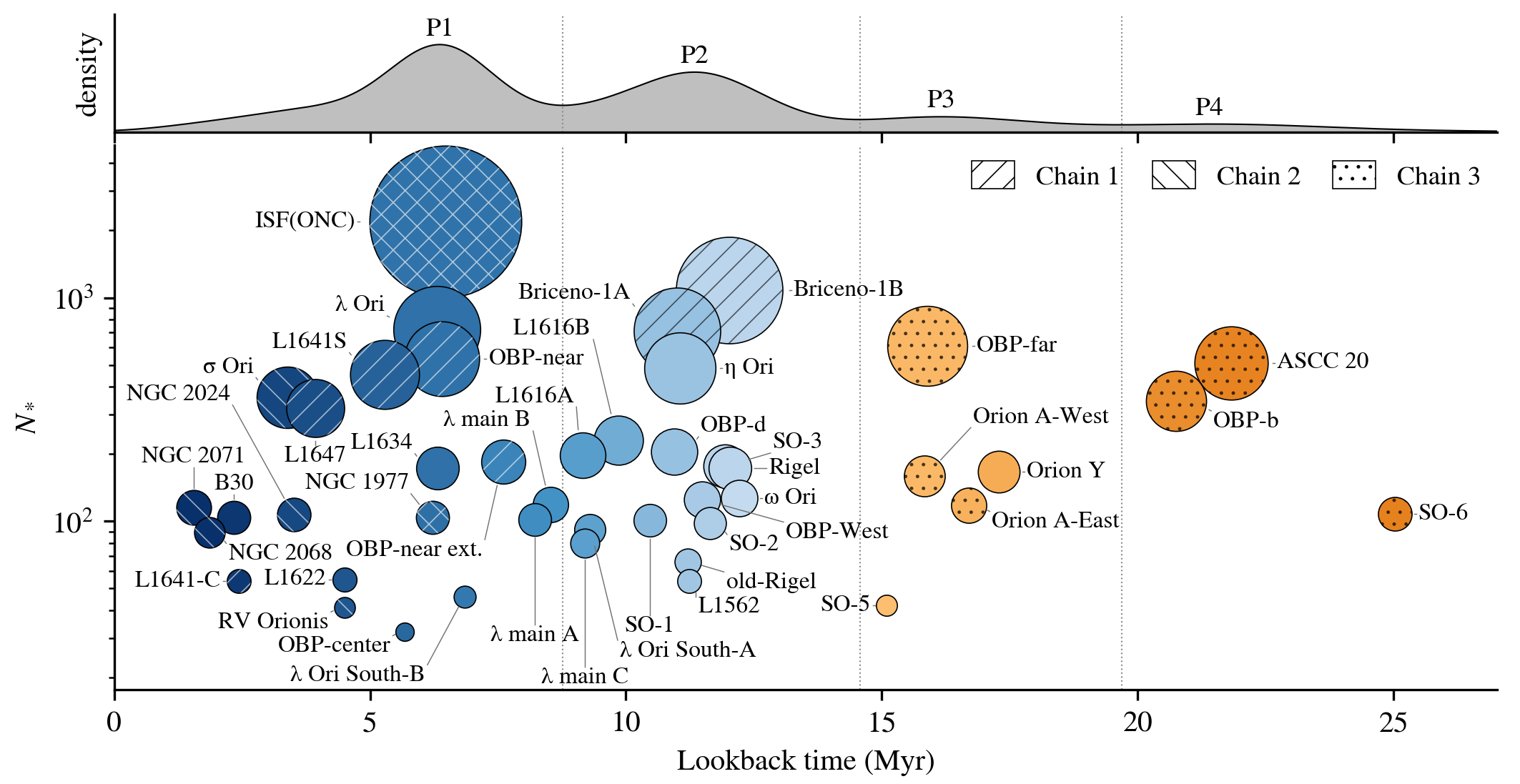}
     \caption{Cluster age versus member number $N_*$ for the Orion clusters ($N_* > 15$). Clusters are color-coded by age and scaled by cluster size, and members of the three proposed cluster chains are highlighted with a pattern. The top of the figure shows the KDE version.}
     \label{fig:formation-timeline}
\end{figure*}

Figure~\ref{fig:age-kde-raw} shows the cluster age distribution using the \citetalias{Rottensteiner_2026} age uncertainties without the 1~Myr uncertainty floor. The positions of $P_1$-$P_4$ of Fig.~\ref{fig:age-kde} remain unchanged to within 0.04~Myr for both panels, and the peaks themselves are narrower. Two additional local maxima appear at $\sim$1.7~Myr and $\sim$3.5~Myr, and contain 4 clusters (3\% of all stars) and 3 clusters (6.6\% of all stars), respectively. However, they cannot be recovered when using uncertainties that better reflect the age scatter from bootstrapping (see Sect.~\ref{sec:Data}). One or both peaks could become stronger when accounting for embedded YSOs not in the \emph{Gaia}-based census of \citetalias{Rottensteiner_2026}, but as we do not have age estimates for these stars, we cannot validate this hypothesis at present. We therefore do not interpret these features and only show Fig.~\ref{fig:age-kde-raw} for completeness. 

The result also does not depend on the exact value of $\sigma_{\rm floor}$. To show this, we test an uncertainty floor range $\sigma_{\rm floor} \in [0,2]$~Myr, with a step size of 0.25~Myr. The upper values act as a control, as they lie above the theoretical merging limit of the features: Two Gaussian components merge into a single maximum once their separation $d$ falls below a multiple of their width, $d < C\cdot\sigma_{\rm peak}$, where $C$ depends only on the ratio of the two peak heights. For peaks of equal height $C = 2$, and for unequal heights $C > 2$ \citep{Felinger_1997}. The maxima of Fig.~\ref{fig:age-kde} are separated by $d \simeq 5$~Myr, and the most unequal pair, $P_2-P_3$ in the star panel, has a height ratio of $0.25$ and $d=4.8$~Myr, corresponding to $C = 2.98$. They are therefore expected to merge for $\sigma_{\rm peak} \gtrsim 1.6$~Myr. Since $\sigma_{\rm peak} > \sigma_{\rm floor}$, the features should merge at a floor at or below $1.6$~Myr. In Fig.~\ref{fig:sweep} we show the age distributions for all $\sigma_{\rm floor}$ values, where each row is normalized to its own maximum. $P_1-P_4$ are present up to $\sigma_{\rm floor} =1.5 $~Myr in both panels and their positions move by $\leq0.4$~Myr in the star panel. In the cluster panel, $P_1$ starts to move $\sim0.8$~Myr towards $P_2$ between $1.25$ and $1.5$~Myr floor as the two peaks begin to merge. At $\sigma_{\rm floor} = 1.75$~Myr, the star panel loses $P_3$ to $P_2$, the pair identified above as the first to merge, and the cluster panel collapses to a single maximum. The two additional young features disappear in the star panel and merge in the cluster panel at $0.75$~Myr, below the adopted floor. 

Lastly, the formation timeline in Fig.~\ref{fig:formation-timeline} shows that the peaked structure of the SFH remains visible without any KDE.

\section{Statistical tests of the SFH structure}
\label{appendix:stats}
We perform the same statistical tests as \citet{Ratzenboeck_2023b} did for Sco-Cen. Localizing the distribution maxima with an excess-mass test \citep{Mueller_1991} and allowing for two modes isolates the intervals 6.2--6.5 and 10.9--12.3~Myr in the star distribution. Together, they hold $58.6\%$ of all members. The cluster distribution yields almost identical intervals containing 15 of the 47 groups. The significance of these modes is given by the dip test \citep{Hartigan_1985}. We calibrate it against $2\times10^{4}$ uniform Monte Carlo samples, which for the star distribution carry the observed cluster weights in random order. The uniform distribution is the least favorable unimodal case, making this calibration conservative \citep{Cheng_1998}. Therefore, we adopt a significance level of 0.1. The test rejects unimodality for the star distribution, with $D = 0.091$ and $p = 0.064$, but not for the clusters ($D = 0.034$, $p = 0.77$). Since each star inherits the age of its cluster, the effective sample size of the star distribution is only 15.6. We also fit Gaussian mixture models with $K \in [1,6]$ components that account for individual age uncertainties and select the best $K$ via the Bayesian information criterion (BIC), computed using the effective sample size. Both panels prefer $K=1$. We can explain this outcome by noting that our effective sample sizes of 15.6 (stars) and 47 (clusters), combined with kernel widths of 1--4~Myr, deconvolve almost the entire observed width into the error term. 

The bootstrap refits confirm the stability of $P_1$ and $P_2$ in the star panel: both are recovered in all five resamples, within 6.3--6.7 and 10.7--11.6~Myr (Table~\ref{tab:peaks}), and their separation stays between 4.4 and 5.2~Myr. $P_3$ and $P_4$ are recovered in only one and three resamples, respectively. The cluster panel, in which each cluster carries the same weight, is more sensitive to individual ages. $P_4$ disappears when the sampler seed changes because one of its three clusters has a bimodal age posterior that can switch between the \citetalias{Rottensteiner_2026} mode (20.8~Myr) and 17.1~Myr. Similarly, the shallow minimum between $P_1$ and $P_2$ (Sect.~\ref{sec:Results}) lets a single cluster move $P_1$ (4.9--6.8~Myr) or merge $P_2$ into it in one of the five resamples. We highlight that the star panel is the more physically meaningful distribution.

Lastly, we investigate the $\sim5$~Myr peak separation by comparing the coefficient of variation for the gaps ($3.5\%$ for the stars and $0.9\%$ for the clusters) to a smooth star formation history with the same uncertainty and cluster number, and using the same kernel. We test 1) the critical bandwidth unimodal density that is closest to our sample \citep{Silverman_1981}, and 2) a uniform distribution. Less than $0.5\%$ of $10^4$ samples can produce at least three gaps of comparable regularity in either case, meaning unimodal star formation histories cannot reproduce the observed regularity. We stress that this is a statement about the four peaks of Fig.~\ref{fig:age-kde}, not a claim that four distinct episodes definitely occurred.

\section{Auxiliary figures showing the physical correlation of the peaks}
\label{appendix:physical-peaks}

\begin{figure*}[ht]
    \centering
    \includegraphics[width=\linewidth]{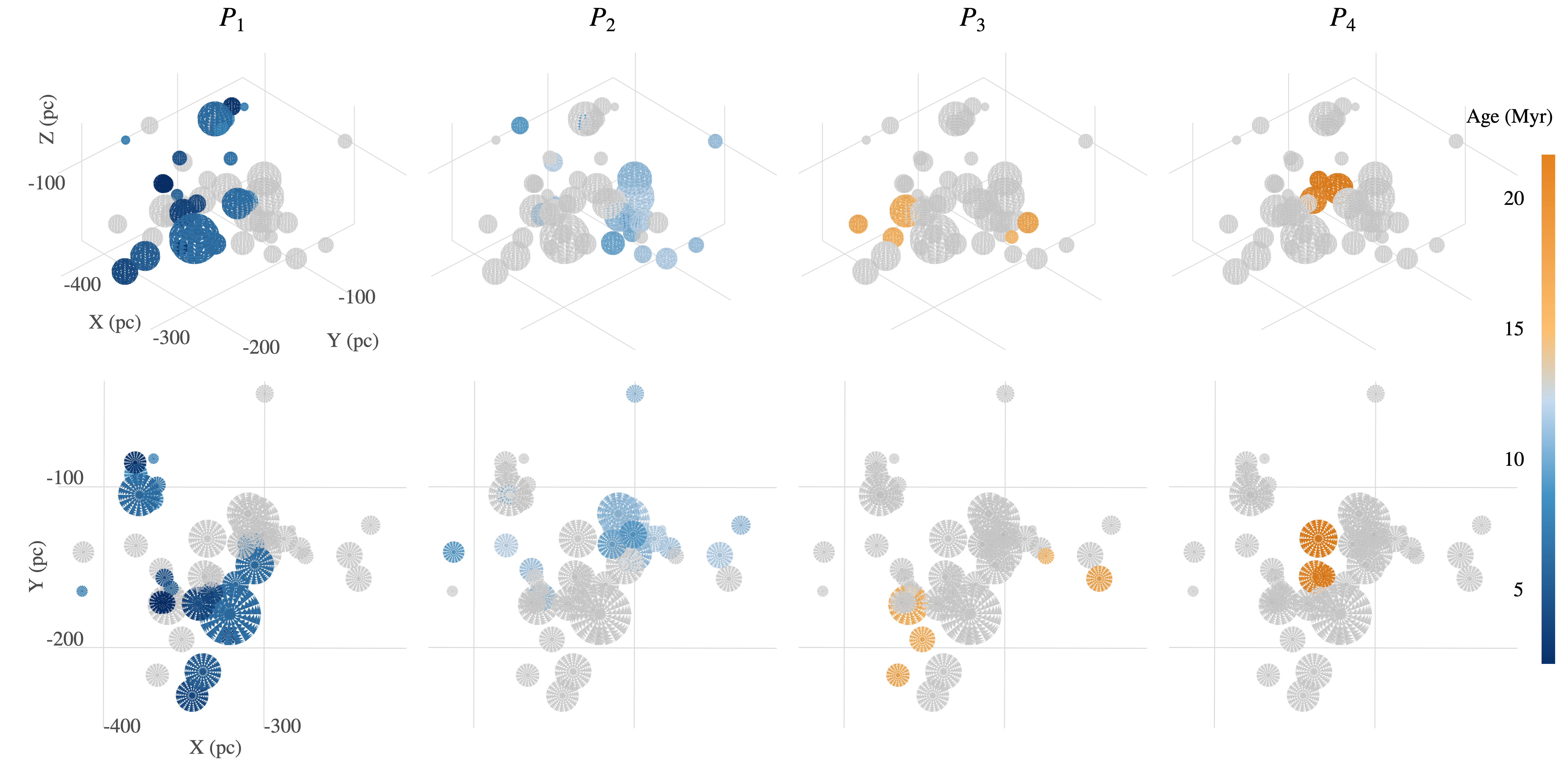}
 \caption{Spatial distribution of the Orion clusters, shown as spheres with radius $\propto N^{1/3}$ at their median positions, separately shown for the four star formation peaks. Clusters not assigned to the respective peak are shown in gray.}
    \label{fig:SFH-spatial}
\end{figure*}

 \begin{figure*}
     \centering
     \includegraphics[width=\linewidth]{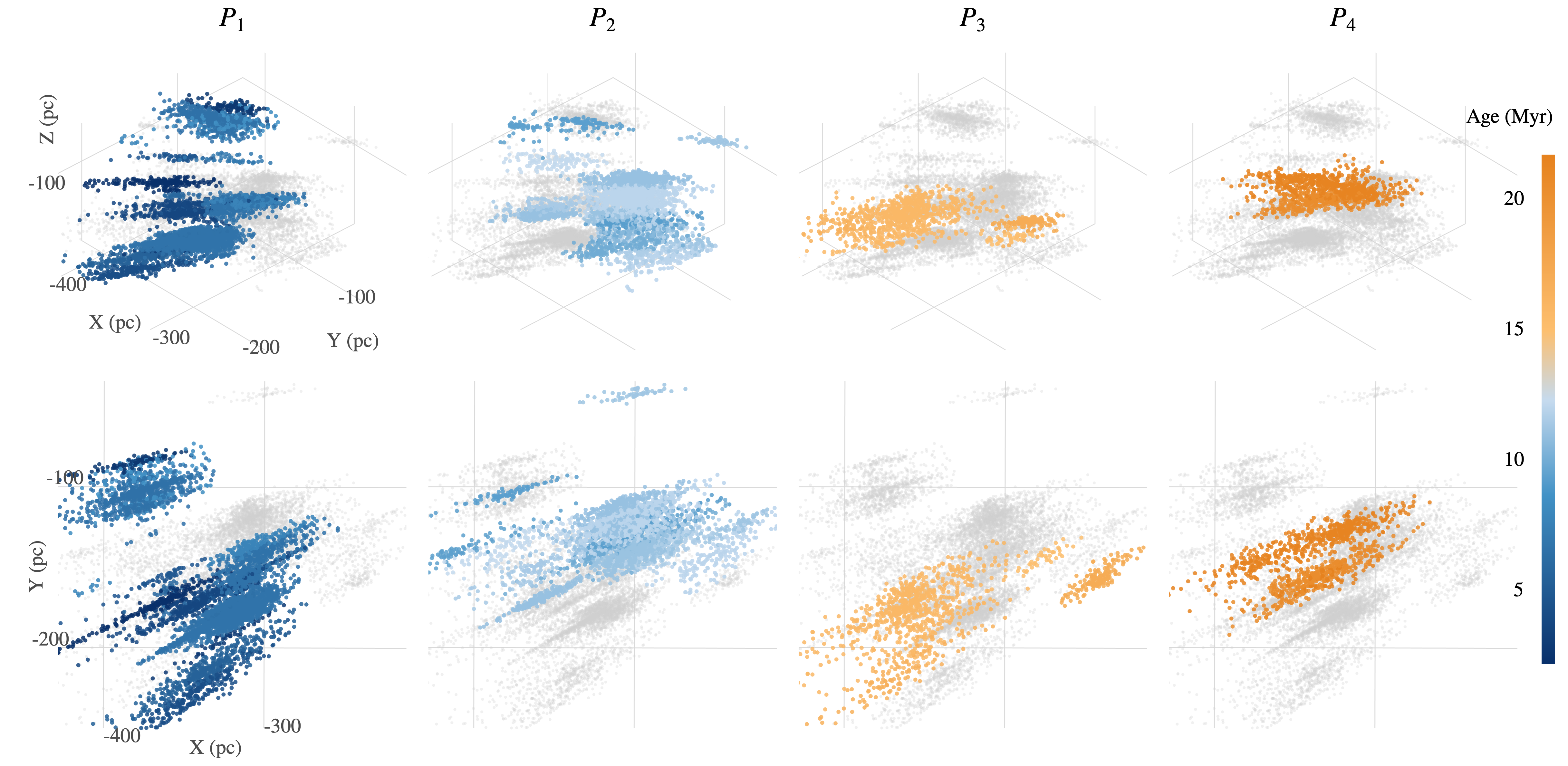}
     \caption{Similar to Fig.~\ref{fig:SFH-spatial}, but showing the 3D distribution of the cluster stars.}
     \label{fig:3D-scatter}
 \end{figure*}

Figures~\ref{fig:SFH-spatial} and~\ref{fig:3D-scatter} show the clusters belonging to each peak, using spheres located at the median cluster position, as well as the cluster members. In the latter case, the clusters are significantly elongated along the line of sight due to the increasing \emph{Gaia} parallax errors \citep[see ][]{Rottensteiner_2026}.

\end{document}